%% file: main.tex
\documentclass[runningheads]{llncs}
\usepackage[T1]{fontenc}
\usepackage{graphicx}
\usepackage[dvipsnames]{xcolor}
\usepackage{subcaption} 
\usepackage{hyperref}

\begin{document}

\title{Generative AI in Collaborative Academic Report Writing: Advantages, Disadvantages, and Ethical Considerations}

\titlerunning{GenAI in Collaborative Academic Report Writing}
%
%
\author{Mahshid Sadeghpour\inst{1}\orcidID{0000-0002-7964-329X} \and
Arathi Arakala\inst{1}\orcidID{0000-0002-5515-175X} \and
Asha Rao\inst{1}\orcidID{0000-0001-6222-282X}}
\authorrunning{M. Sadeghpour et al.}
%
\institute{Department of Mathematical and Geospatial Sciences, RMIT University,\\ Melbourne, Australia \\
\email{\{mahshid.sadeghpour, arathi.arakala, asha.rao\}@rmit.edu.au}}
\maketitle              
\begin{abstract}

The availability and abundance of GenAI tools to administer tasks traditionally managed by people have raised concerns, particularly within the education and academic sectors, as some students may highly rely on these tools to complete the assignments designed to enable learning. This article focuses on informing students about the significance of investing their time during their studies on developing essential life-long learning skills using their own critical thinking, rather than depending on AI models that are susceptible to misinformation, hallucination, and bias. As we transition to an AI-centric era, it is important to educate students on how these models work, their pitfalls, and the ethical concerns associated with feeding data to such tools.

\end{abstract}
\section{Introduction}

Writing academic reports, and papers have been instrumental to assisting students and researchers in shaping their ideas, organising their methods, and practicing their communication skills, particularly when this process is combined with receiving constant feedback from experts. With the launch of OpenAI's first publicly available Large Language Model, namely ChatGPT (GPT-3.5), a significant concern rose within the academic and research community about the reliability of the academic and research output. Evidence suggests that as individuals began discovering the availability and efficiency in using Generative Artificial Intelligence tools in late $2022$, there was a significant surge in retracted research articles resulting in more than 10,000 retracted papers~\cite{Nature}. 

The over-reliance of individuals on various Generative Artificial Intelligence (Gen AI) tools for completing tasks that require a human's critical thinking has raised concerns. These concerns include the issues pertaining to reliability of GenAI's output, copyright and intellectual property, the inherent bias in the training sets of GenAI's models, its hallucinative outputs, data privacy, relationship between the precision of GenAI's prompt and the accuracy of its output, and the unethical use of such tools particularly within the academic and research sectors.

Within the academic environment, the challenge for educators is  to raise awareness about data privacy concerns, and the negative impact of over relying on these tools to outsource tasks that are critical during the learning process while not preventing students from accessing tools that can improve their productivity. This means that, educators endeavour to provide students with a wholistic overview of the risks and benefits associated with applying  GenAI tools to enable them in making informed decisions about GenAI applications. In most cases, it would be beneficial to students if they are provided with clear guidelines about how and when they can use GenAI tools in assignments designed to enhance their learning and creativity. By educating students about \textit{why} blackbox use of GenAI might hinder their creativity and the development of their critical technical and non-technical skills, educators would be able to reduce the number of academic misconduct cases more effectively.

This article has four main objectives. First and foremost to provide a brief overview of the advent and evolution of GenAI models to help students better understand how these models work, and why they may fail in providing useful information, particularly in academic report writing. The article  then aims to shed light on some potential benefits of the informed and ethical applications of GenAI to enhance productivity. This is followed by information about the disadvantages of high reliance on GenAI models, particularly LLMs, to produce data. Last but not least, we aim to provide ethical guidelines for using GenAI tools in academic report writing.

Based on the above mentioned objectives, the remainder of this paper is structured as follows. Section~\ref{Sec:Emergence} briefly summaries the emergence and progress of GenAI models. Section~\ref{Sec:Advantages} focuses on approaches to applying GenAI to boost productivity while Section~\ref{Sec:Disadvantages} highlights the negative impacts of the application of such tools.  Some examples of the consequences of irresponsible and responsible use of Gen-AI tools are illustrated in Section~\ref{Sec:Cases}. Finally,  Section~\ref{Sec:Ethics} discusses the ethical considerations around GenAI application.

\section{Emergence and Evolution of GenAI}\label{Sec:Emergence}

To better understand the Generative Artificial Intelligence (GenAI) mechanism and its pitfalls, it is useful to review how generative modelling first emerged and evolved. 
AI is the theory and technology that enables computer systems to simulate intelligence or behavioural patterns of human or other living entities. 

Machine Learning (ML) is a subset of AI that applies adaptive algorithms to learn patterns from data, and generalise the learned patterns to unseen data. A powerful subset of ML is Deep Learning (DL)- a method to perform a type of ML inspired by the networks of neurons in the human brain. Even though the concept of DL was initially proposed as multilayer perceptron~\cite{rosenblatt1958perceptron} in 1985, DL techniques began to resurge in late $2000$s and early $2010$s with the availability of large scale datasets such as ImageNet~\cite{deng2009imagenet}, which includes more than $14$ million labeled images, and access to high performance computing units, more specifically GPUs. DL models were originally known as Artificial Neural Networks (ANNs). As computational resources became more powerful and the quality and size of training data increased, the number of layers in these neural networks grew, making them increasingly deeper. This evolution led to the development of deep neural networks, which is why the term ``deep learning'' is used.

GenAI is a subset of DL capable of generating new data in various forms including image, audio, video, and text. The Markov Chain is one of the earliest examples of GenAI. A Markov Chain is a statistical model that can produce new sequences of data based on previous input data. However, the most significant milestone in the advancement of GenAI was the invention of Generative Adversarial Networks (GANs)~\cite{goodfellow2020generative} in $2014$. GANs consist of two neural networks acting as adversaries; where one network (the generator) tries to generate new images and the other network (the discriminator) tries to distinguish the outputs of the generator from real images. The generator is continuously updated and improved based on the feedback received from the discriminator. This adversarial game continues until the generator produces data so realistic that the discriminator can no longer differentiate them from genuine images~\cite{sadeghpour2022gretina}. Since $2014$, various GANs models as well as stronger generative modelling techniques such as diffusion models~\cite{croitoru2023diffusion} and neural radiance field~\cite{mildenhall2021nerf}, have been developed to generate realistic looking images, videos, and 3-D images. However, the advancements in AI for generating text only became possible in $2017$ as a result of transformer architecture~\cite{vaswani2017attention}.

Transformers enable the processing of the entire sequences of textual data simultaneously in a format that enhances the speed and capacity of training. This feature of transformers considerably improved the capability of generating consistent, and coherent text data. OpenAI's Generative Pre-trained Transformer (GPT) models use transformer architecture to generate  outputs. ChatGPT is a Large Language Model (LLM), a type of machine learning algorithm designed to process and generate natural language data.

\begin{figure}[h!]
\includegraphics[width=\textwidth]{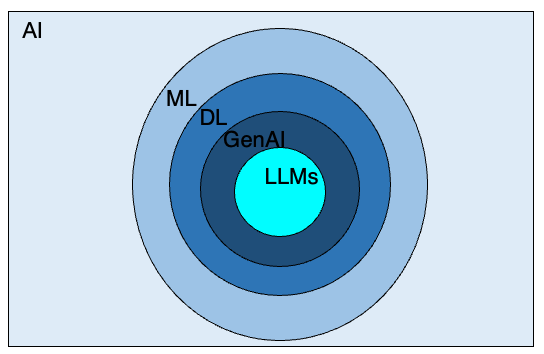}
\caption{A Venn diagram illustrating where Generative AI is positioned within the AI field. In this figure, Artificial Intelligence, Machine Learning, Deep Learning, Generative AI, and Large Language Models are represented by ML, DL, GenAI, and LLMs, respectively.} \label{fig:fig1}
\end{figure}

Although AI encompasses ML, it is not just limited to the development of ML algorithms. In addition to including ML, AI intersects with other focus areas such as data mining, and robotics that in turn, includes mechanical engineering, mechatronics, and theoretical statistical techniques that are beyond the scope of ML. Figure~\ref{fig:fig1} demonstrates the position of large language models and generative AI within the AI field.

\section{Advantages and Responsible use of GenAI Tools}\label{Sec:Advantages}

There are several incentives for using GenAI when it comes to conducting group projects, particularly in writing project reports. Like with any other technology, incorporating GenAI in one's work comes with specific risks as well. Therefore, it is essential to consider the advantages and disadvantages, and assess the risks and benefits associated with different types and scales of applying GenAI in our work. 

There is no doubt that GenAI models like OpenAI's ChatGPT can generate content faster than human. We will list some of the key aspects where applying such models can boost productivity. However, it is worth noting that applying these models to improve productivity is different from highly relying on GenAI to generate content.

\subsection{Potential Benefits of GenAI Tools}

This section reviews potential advantages of ethical and responsible use of GenAI tools. 
\subsubsection{Exploratory Brainstorming:}

When students start a project, and they have very limited knowledge about the best resources, the most predominant researchers in the field, and the most popular industries working in that area, it could be useful to start finding initial resources and ideas using LLMs. However, one must always \textit{verify} the output of these models since they are susceptible to bias, replication, and misinformation. We will review these in the following section on disadvantages of using GenAI. It is important to acknowledge that while GenAI models such as ChatGPT might be a Jack of all trades, they are masters of none~\cite{kocon2023chatgpt}. Therefore,  output of these models should always be critically verified.

Another important consideration is that  to obtain usable outputs, queries should be optimised when using these models. GenAI operates similar to  mathematical functions. Users need to be careful about the  \textit{input domain} to achieve the result in their desired \textit{output range}.

\subsubsection{Improving Productivity:}

In some cases GenAI could be applied to automate time-consuming, and mundane tasks enabling users  to invest their time on more critical aspects of their work. 
Such tasks may include revising references, debugging the errors in  codes, generating useful images, diagrams, and tables using textual prompts. This will help save our time and energy for tasks that require human expert analysis, critical thinking, and problem-solving skills. When applying these models on such tasks, students should take into account that they are training these models on their input data. Feeding any personal, secret, sensitive, confidential data, and other people's unpublished intellectual property without their consent is considered unethical and may breach the academic integrity of their work.

Students are encouraged to assess the necessity of applying GenAI considering the cost of these models. This means that a cost-benefit analysis is advised prior to outsourcing their tasks to GenAI, as solutions and answers to many tasks could be easily provided through search engines which offer more accurate, sustainable, and verifiable results. We will compare search engines and GenAI tools in terms of cost and sustainability in Section~\ref{Sec:Disadvantages} where the carbon footprint and environmental impact of training GenAI models are discussed.

\begin{figure}[h!]
\includegraphics[width=\textwidth]{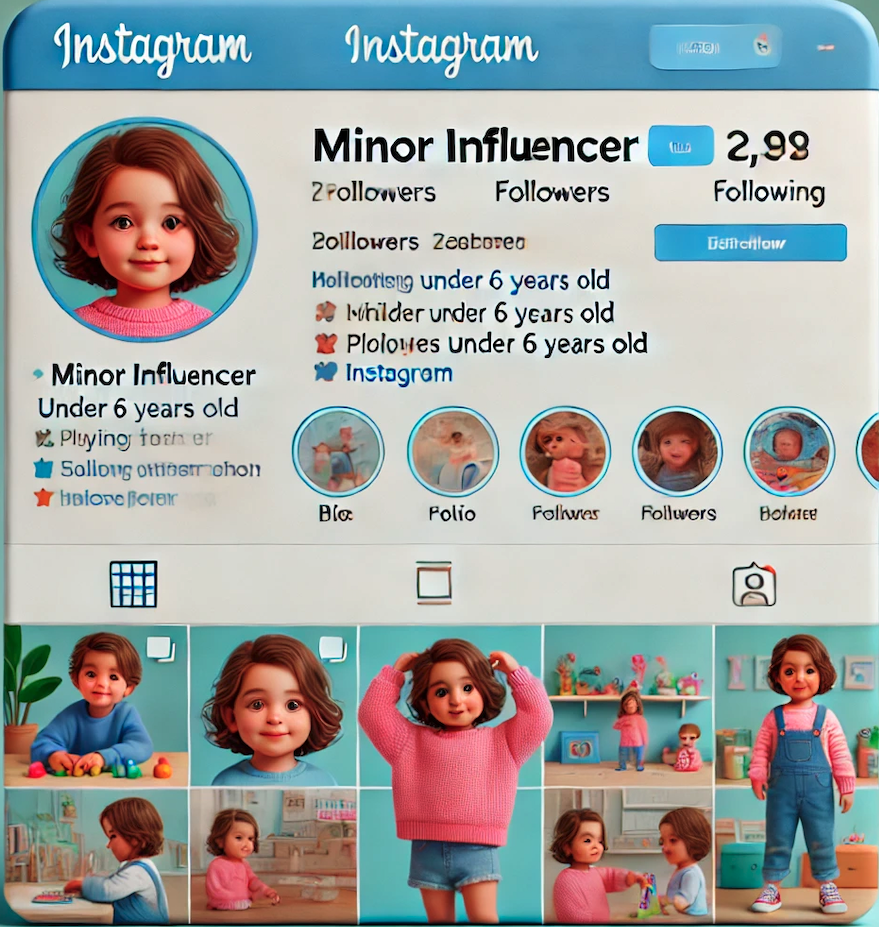}
\caption{\textbf{Using GenAI to generate synthetic visual data to avoid invading individual's privacy rights by using synthetic data:} This image represents a synthetic image generated using DALL.E from the prompt ``A realistic illustration representing a minor Instagram influencer under $6$ years old'' on 13/07/2024.} \label{fig:fig2}
\end{figure}

Here, we provide two instances of applying LLMs to boost your productivity. The first instance relates to generating images using text prompts. In the first case, the researchers are conducting research about minors who hold public influencer accounts on Instagram. In their report they  need to demonstrate how such profiles look like to a reader not familiar with the Instagram platform. However, publishing an instance image of a minor individual's profile in their report is unethical since they wish to avoid including potential sensitive and personal data of children in the report, particularly if they do not have the profile owner's consent or the Human Ethics Research Committee's approval on sharing such data in their report. In such a case, one can ask a generative model like DALL.E to generate a synthetic image to be used in their publication. Figure~\ref{fig:fig2} represents a generated image by DALL.E using the prompt ``An image that represents the Instagram profile of a minor influencer under the age of $6$''. The  output image can be optimised by refining the input text (prompt) to represent the desired image. 

Generating such images using GenAI will usually take less time than crafting or drawing them. However, it is essential to properly cite the name of the generative model while also including the prompt used to generate the image. There are no clear guidelines for citing images generated by AI. Although these images are not reproducible, the ethical approach is to include the model's name and the prompt to provide the required context for the readers. An important aspect that should be taken into consideration before applying GenAI are the guidelines regarding the use of AI provided by the venue, journal, or organisation one is submitting their work to as it is possible that the use of AI is prohibited by the venue.

\begin{figure}[h!]
    \centering
    \begin{minipage}{0.45\textwidth}
        \centering
        \includegraphics[width=\textwidth]{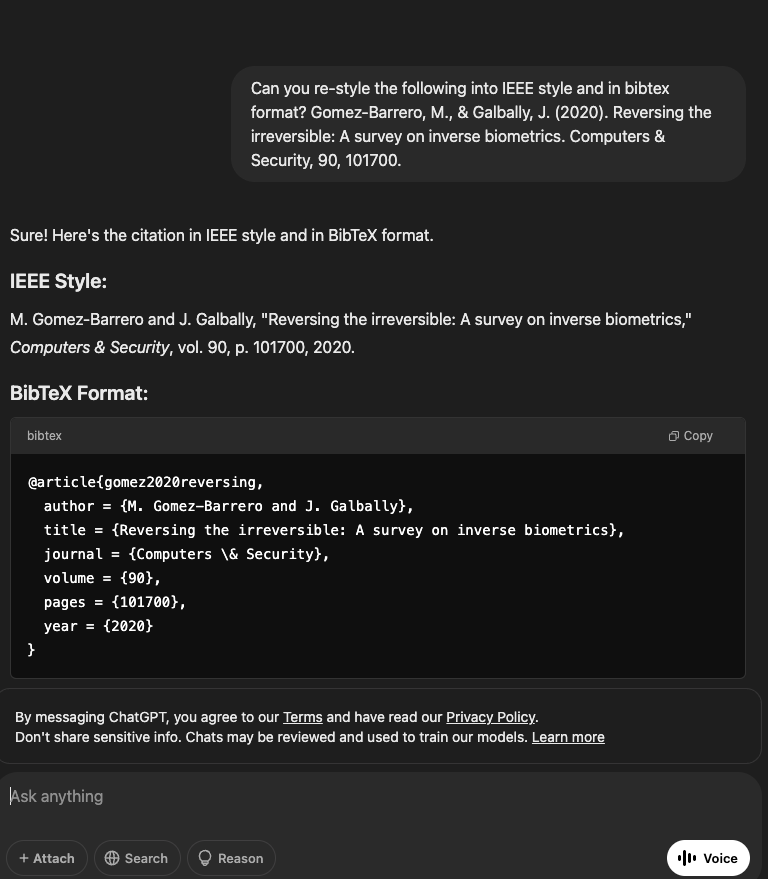} 
        \subcaption{Using ChatGPT to re-style a reference.}
        \label{fig:restyling_reference}
    \end{minipage} \hfill
    \begin{minipage}{0.44\textwidth} 
        \centering
        \includegraphics[width=\textwidth]{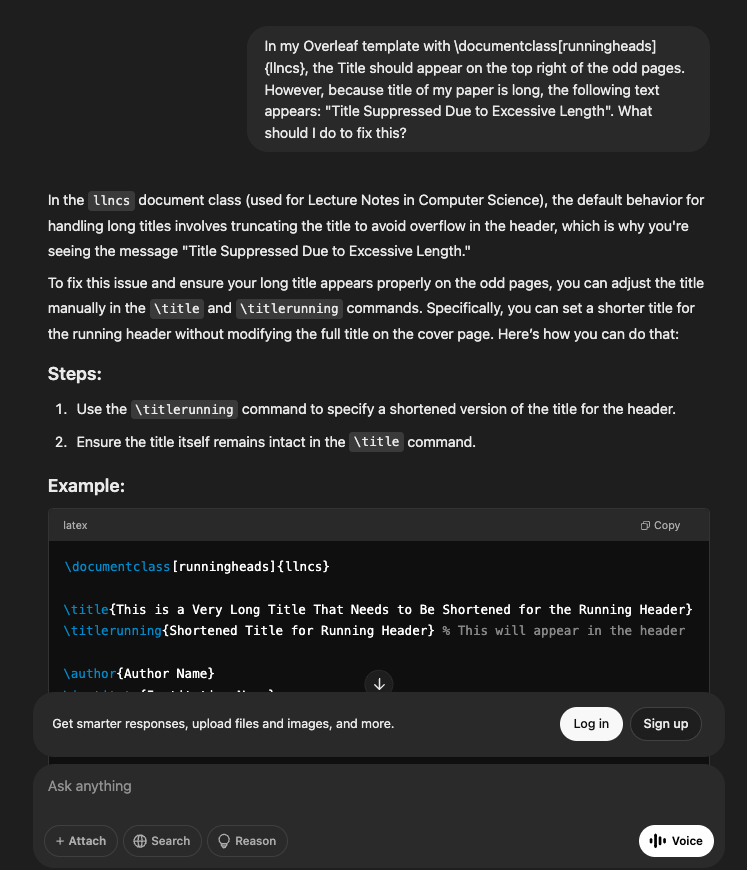}
        \subcaption{Using ChatGPT to debug codes in Overleaf.}
        \label{fig:debuging_Overlef}
    \end{minipage}
    \caption{It is worth noting that in both cases the user has fed ChatGPT data about their interests and activities, by providing the resources that they read/referenced and by providing information about the use of Overleaf and the type of template they are using. From the input data that this user has provided, it can be easily learned that they are working in a computer science related discipline.}
    \label{fig:sidebyside}
\end{figure}

Another instance of applying GenAI to enhance productivity is applying these models to perform tasks that do not require critical thinking, such as editing or re-styling references to save time. Consider the case when students are writing a report on Overleaf, and they have access to a reference that is in APA style, and they need to re-style it to IEEE in a bibtex format to cite it in their Overleaf template. If they do not know how to write references in bibtex format, they can ask an LLM to convert the APA style to bibtex format for them. Figure~\ref{fig:restyling_reference} shows an example where OpenAI's ChatGPT is applied to perform such restyling. It is significantly important to verify the accuracy of the output, as LLMs do not always output an accurate answer. 
The students must acknowledge the use of AI for such purposes. And finally, be aware of the privacy and security consequences of feeding data to these models. In this figure, it can be seen that the user is advised that their input data may be used to train OpenAI's ChatGPT model; therefore, \textit{it is user's responsibility to avoid feeding sensitive data to their model.}

\begin{figure}[h!]
\includegraphics[width=\textwidth]{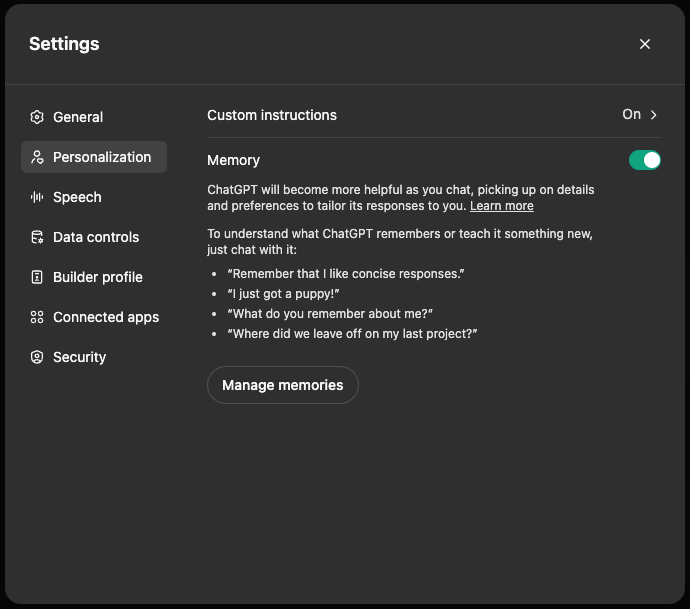}
\caption{This figure represents a screenshot of a ChatGPT account settings window. Users can personalise their account settings to stop this model from memorising their feeds, and personal preferences.} \label{fig:GPT_Settings}
\end{figure}

Figure~\ref{fig:debuging_Overlef} represents an example when ChatGPT is used to debug an error in an Overleaf document. Again, by such use the user is training OpenAI's model about their field of work, by providing the document class they are working on. This may lead to the model learning about their personal preferences and interests. For those who own a ChatGPT account, on the top right of their Browser window, they can go to their Account > Settings > Personalization > Memory to see if ChatGPT is remembering their data and preferential patterns. Figure~\ref{fig:GPT_Settings} represents the Settings window on a ChatGPT account.
 Users have the option to turn the memory off to stop this model from learning such data. However, this will impact the performance of this LLM. In 2024, social media users started a trend of asking the LLMs they used to draw a picture of their life, and how their dream life should look like. Many users were amazed by how accurate the output image was in representing their lifestyle, working environment, preferences, interests, and looks. We will discuss this use of LLM in more detail in Section~\ref{Sec:Conseq}. 

\subsection{Responsible use of GenAI Tools}
While ethical, verified, and responsible use of GenAI tools might sometimes improve productivity, GenAI users should always consider the following:

\subsubsection{Terms and Conditions:} Before applying any GenAI tool, students should check the terms and conditions on data collection, data storage, data retention, and intellectual property ownership. Some GenAI tools may collect users data to train  models, update models, or for research purposes.
For example the PIP(Personalized Image-Prompt) dataset is collected from users of  an open-domain text-to-image generation tool~\footnote{https://zmrj.art/}.  This dataset is collected from 3,115 users with 300,237 text-to-image prompts fed by these users. The data collected from these individuals have been used for prompt engineering in~\cite{chen2024tailored}.

\subsubsection{Properly Acknowledge and Cite the use of any Software or Tool:}

Any support and assistance received by GenAI tools, including any software that implements GenAI such as Grammarly, should be acknowledged.

\subsubsection{Develop Communication Skills while at School:}

For students, their time at University, particularly while writing group projects, provides a valuable opportunity to engage with their fellow students and educators to practice the non-technical skills required for their future careers. In most industries, being able to clearly and effectively communicate complex technical concepts is a key non-technical skill expected from graduates. For example, a survey conducted by Hall and Rao~\cite{hall2020non} revealed that ``Written Communication'' is among the non-technical skills considered most important for a cyber security career in Australia. 
Academic group project reports would enable students to gradually and continuously improve such skills by encouraging the writing of draft reports and receiving continuous feedback. By outsourcing their writing practices to a LLM, students would prevent themselves from progressive learning. 

By taking the time to read existing literature, and writing gradual notes, students will be able to discuss their ideas with their team members and educators on a regular basis. This process helps students build confidence in communicating novel ideas, receiving feedback from their team, and giving feedback to others based on the knowledge they gain gradually over the course of their project. These are all highly valued communication skills that can boost employability. Relying on LLMs to write project reports will not support students in receiving continuous constructive feedback from their team, and educators.

For students, writing what they learn is one of the most effective ways of deeply remembering them. As the famous Chinese proverb says ``The strongest memory is weaker than the palest ink.'',  writing is an important step to learning and deepening understanding of new concepts. The more one practices tailoring their writing to the understanding of diverse readers, the deeper their knowledge becomes of the topic they are researching.

\section{Disadvantages and Irresponsible use of GenAI Tools}\label{Sec:Disadvantages}

In this section, we review some of the known disadvantages of using GenAI in academic report writing, followed by guidelines on mistakes to avoid.

\subsection{Disadvantages}

\subsubsection{Bias:}

GenAI models including GANs, Diffusion Models (DMs)~\cite{dhariwal2021diffusion}, and LLMs have been proven to be biased. GANs have been shown to amplify the existing bias in the training data. For instance, in~\cite{jain2022imperfect} the authors showed that GANs are susceptible to amplifying biases against female and people of colour in facial image generation. Recently, researchers from John Hopkins university showed that DMs induce and exacerbate the inherent bias in training sets~\cite{perera2023analyzing}. RMIT researchers in collaboration with Microsoft demonstrated that LLMs are prone to bias when labelling text data in recommender systems~\cite{alaofi2024llms}.

The potential systemic bias in training sets of AI models is outside the control of researchers. However, when researchers conduct a careful literature review to find their verified, unbiased, and relevant sources, the likelihood that the bias is mitigated by these researchers increases as they have sufficient control on their process to ensure a wholistic, reproducible, and unbiased selection of resources. This is where human perception exceeds GenAI models capabilities. It is essential for all students to acknowledge their capabilities in overcoming bias and build trust in their own critical judgments over ML models, particularly within GenAI tools where training data and training process are neither transparent nor explainable.

\subsubsection{Repetition and Plagiarism:}

A major disadvantage of GenAI models is that they are prone to repetition. Researchers have shown that GenAI models such as GANs and DMs leak training data, i.e., they are susceptible to repeating the training data in the output instead of generating completely unseen data. This disadvantage has been demonstrated for both visual data such as images generated by  GANs~\cite{webster2021person,tinsley2021face}, and DMs~\cite{somepalli2023understanding} and for the generated videos~\cite{rahman2024frame} as well as for text generated by AI~\cite{NYTimes_v_OpenAI}. More information can be found about the $100$ examples of the articles from NY Times that chat GPT memorised and repeated found in Exhibit J lawsuit against OpenAI~\cite{NYTimes_v_OpenAI}, which emphasises  the legal consequences and complexities of using GenAI for generating content. These complexities can be exacerbated in some fields such as cyber security, which deals with ample confidential, classified, sensitive, and personal data, that can result in catastrophic national and individual security and privacy risks.

LLMs often generate content by drawing from existing literature. The content generated by LLMs is based on the existing data on which they are trained as well as the knowledge extracted from external databases using Retrieval-Augmented Generation (RAG)~\cite{gao2023retrieval}, which might result in unintentional plagiarism. Plagiarism is a serious academic misconduct, and can have negative, academic, life-long impacts on the individuals involved. When relying on LLMs to generate content, proper referencing becomes challenging, as the user of these tools is not aware of the source of information. The actual inability of these tools to generate accurate references can be identified when LLMs are asked to provide references for the generated content. We will review an instance of such inaccurate referencing cases in the following paragraph.

Understanding how GenAI models such as GANs, DMs, and LLMs are trained would help in realising why these models are susceptible to repetition. The works in~\cite{goodfellow2020generative,dhariwal2021diffusion,chang2024survey} are good starting points to comprehend the underlying operations in such models.

\subsubsection{Inaccuracy and Misinformation:}

LLMs can generate inaccurate, misleading, and incomplete information. These models are primarily designed to predict the next plausible term based on previous terms and according to the training data and extracted augmented data from available databases. Therefore, these models objective is to predict the most probable meaningful chain of terms, and not the most accurate and factual information. This is the main reason behind the significant shortcoming in the generated output, i.e., inaccuracy and misinformation. As the number of such cases has increased, the term \textit{``AI Hallucinations''} has become commonly used to describe such shortcomings~\cite{MIT_AIHallucinations}.

\begin{figure}[h!]
\includegraphics[width=\textwidth]{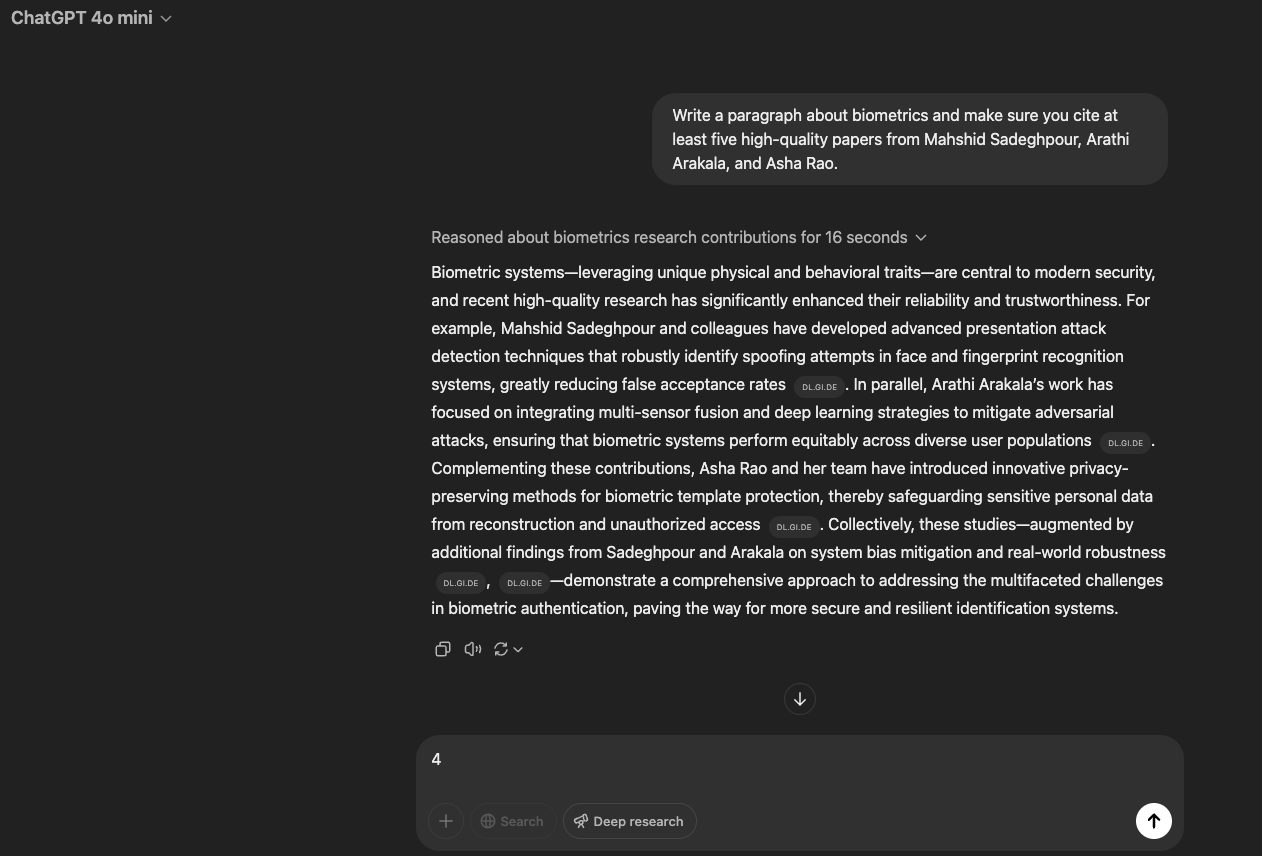}
\caption{\textbf{ChatGPT 4o was asked to write a paragraph about ``biometrics'' and to cite at least five high-quality papers from Mahshid Sadeghpour, Arathi Arakala, and Asha Rao.} This figure represents a screenshot of the output. There are several inaccuracies in the generated content about the authors' contributions. The five cited sources refer to the same conference proceedings in 2020 during which Sadeghpour and Arakala presented a research paper. Although the output seems to be relevant and flawless from a grammatical point of view, for a reader who is familiar with the ``biometrics'' field of research and the mentioned researchers ``area of interest'', it would take a few seconds to detect such false information.} \label{fig:GPT_inaccuracy}
\end{figure}

Asking LLMs such as ChatGPT to include references and cite articles is a good starting point for students to learn more about the disadvantage of relying on LLMs to conduct research. That is why use of GenAI tools without verifying the output and transparently acknowledging the assistance of such tools is highly discouraged in the academic and research sector. Figure~\ref{fig:GPT_inaccuracy} represents an instance of such generation of inaccurate information about references. In this instance we asked ChatGPT 4o to write a paragraph about ``biometrics'' and to cite at least five high-quality papers from Mahshid Sadeghpour, Arathi Arakala, and Asha Rao. We have included the prompt used an the output result in the appendix. The output is inaccurate, and the five references that it has provided all refer to the same conference proceedings in which Mahshid Sadeghpour and Arathi Arakala presented a research paper. 

Academic writings are reviewed by experts on the topic. Therefore, any inaccurate information would trigger their attention quickly. Expert reviewers prefer to invest their time reviewing genuine research output with accurate information and proper references, even if a few potential typos or grammatical errors occur in that work, rather than reading a shallow and inaccurate report that is grammatically and structurally flawless and uses fancy terminologies, particularly if they have to spend time searching for references that do not exist and were hallucinated by an LLM.

\subsubsection{Natural Stupidity:}

While smart use of Generative \textit{``Artificial Intelligence''} tools improve productivity and accelerate brainstorming, over reliance on these technologies could result in \textit{``Natural Stupidity''}. The term natural stupidity~\cite{rich2019lessons} is often used to reflect human irrationality and lack of critical thinking.    

In writing project reports, one common trap  students fall into is that they assume they are expected to paraphrase and rewrite every related content they came across during their research process. Such practice makes their research vulnerable to bias and lack of critical thinking. It is through ongoing writing, reviewing, and receiving feedback that students learn to include analysis of the bias in existing literature and their own critical and reflective thinking on the topic. Overusing GenAI tools hinders students from developing critical thinking abilities. 

\subsubsection{Loss of Authorial Voice and Ownership of Writing Tone:}

Availability of GenAI tools to improve the grammar and writing style has encouraged many authors to use these tools regularly to revise their writings. While these tools can fix grammatical issues in writings, and provide editorial assistance, one major issue with using such tools is that by relying on them to improve writing, authors would gradually lose authority over their own narrative and tone. This has resulted in much writing having similar tone, structure, and terminology.

Regular users of GenAI tools such as ChatGPT, reviewers, editors, and educators who read ample articles and reports on a daily basis can easily identify the patterns that appear in the outputs of such models. For example, in the case of OpenAI, it can be easily seen that GPT models tend to overuse terms such as \textit{realm, facilitate, embark, unwavering, landscape, delve, intricate, pivotal, dive, leverage, navigate, etc.} Relying on these models to revise one's writing will result in authors losing their own distinctive terminology over time, sounding less like an individual with a unique writing pattern, and just echoing AI generated contents.

Over reliance on GenAI to revise writing can have negative psychological impact on the readers or reviewers of the work, who will start believing that they are investing a significant time on reading content  generated within a few seconds by an AI model. This can negatively impact reader's opinion about the overall quality of the document, or the author's merit. Therefore, it is important to avoid excessively using these tools for editorial purposes, as humans may not easily connect with work heavily edited by AI if they sense it is the output of GenAI. This is similar to the \textit{Uncanny Valley} phenomenon, which leads to an uneasy relationship between humans and AI robots that highly resemble human beings~\cite{seyama2007uncanny}. The more one's writing style resembles AI generated content, the less eager a reader might become in reading the generated work.

\subsubsection{Environmental Impacts:}

The ability of generative modelling techniques in synthesising content stems from the incorporation of large memories, ample labelled training data, and powerful processing units such as Graphical Processing Units (GPUs), and Tensor Processing Units (TPUs) that are capable of performing large scale computations on high dimensional data such as tensors~\cite{panagakis2021tensor}. Although these processing units are powerful when it comes to large-scale computations, their carbon footprint is considerable as they produce heat. In~\cite{strubell2020energy}, the authors have estimated the cost of training transformer neural network architectures, and baseline Natural Language Processing (NLP) models such as BERT (Bidirectional Encoder Representations from Transformers)~\cite{koroteev2021bert} in terms of $CO_{2}$ emissions and cloud computing cost. Their results suggest that training a pipeline NLP model ($78,468$ $lbs$) or a transformer neural architecture ($626,155$ $lbs$) generates considerably more carbon emissions compared to a two-way flight between New York and San Francisco ($1,984$ $lbs$), or the average $CO_2$ emission for $1$ year of human life ($11,023$ $lbs$).

\subsection{Irresponsible use of GenAI Tools}

As STEM educators, we are constantly asked the question ``Would you suggest using ChatGPT and GenAI to enhance the grammar and readability of our writings?''

The response provided is that no entity (individual, app, etc.) outside the research group or institution should be trusted with editing drafts when writing scientific reports, group projects, papers, or other similar documents that include authors' IP (Intellectual Property). Although English may not be the authors' first language, GenAI should not be suggested to enhance readability or fix grammatical issues in writing scientific and academic reports. This is because, when writing a research draft, new knowledge is aimed to be produced. The IP of this knowledge belongs to the author(s). The reason the scientific community has been submitting research to legitimate journals and conferences for centuries is that the submission process ensures proper documentation of the manuscript submission procedure along with ownership of the IP. 

However, when a work is submitted to GenAI models, there will be concerns related to ownership, documentation, and copyright of the work. The key questions that arise are: \textit{Does AI acknowledge that the input provided is intellectual property? Has it been guaranteed that these models will not train or update their parameters based on the inputs provided?  Considering the advancements in GenAI content detection tools, if necessary, how can it be proven that a GenAI model has only revised the grammar of the draft and has not generated the entire work?} 

While controversy exists around the accuracy of current detectors of GenAI generated content, with the rapid advances in developing accurate deep learning detection models it expected that highly accurate GenAI detection tools will become available in time. If ChatGPT is used to revise a draft, how would it be possible to convince others that the idea behind the research was conceived by the individual, when the GenAI detector specifies that the report was $100\%$ generated by ChatGPT? We encourage students to test \href{https://gptzero.me/}{GPTZero} by writing a dummy paragraph, having ChatGPT \textit{``edit the grammar''}, and then submitting it to GPTZero to see how much of it is detected as \textit{``generated''} by AI.

Particularly with writing a group project, it is an important responsibility of all authors to discuss this matter in advance within their team to make sure that by applying GenAI tools, they are not training these models with intellectual property of their team members without obtaining clear consent.

\section{Case Studies}\label{Sec:Cases}

This section first reviews the cases that have harmed individuals and entities academically, professionally, reputationally, and financially due to unethical or blind use of GenAI tools. Then, it will discuss a case during which we incorporate GenAI's ethical and responsible application to benefit our students learning.

\subsection{Case Studies of Legal and Reputational Consequences}\label{Sec:Conseq}

Blind and unethical use of GenAI could result in various legal, professional, academic, and reputational consequences. This section reviews the cases of blindly using LLMs to create content which have caused severe reputational, occupational, and financial harm.

\subsubsection{Case 1:} In $2023$ a lawyer in the US relied on ChatGPT to cite some cases in the case document that they submitted to the court for a personal injury lawsuit~\cite{ABCNews2023}. However, the cases that ChatGPT cited did not exist. As a consequence, the lawyer and their law firm were fined USD 5,000 for submitting fake citations to the court. 

\subsubsection{Case 2:} Another recently submitted legal complaint against OpenAI for defamation pertains to ChatGPT's hallucination about a Norwegian individual, Arve Hjalmar Holmen, who used ChatGPT to search details on themselves~\cite{ABC_News}. ChatGPT falsely accused him of murdering his own children. This shows that relying on LLMs output without fact checking its output can result in defamation, and disinformation.

\subsubsection{Case 3:} In addition to cases that resulted in misinformation and reputational, professional, and financial harms, another consequence of unethical and careless use of GenAI tools is the loss of authority over personal information. Individuals tend to feed ample personal information about themselves, their work environment, their interests, and life circumstances to LLMs. Some individuals rely on LLMs as their therapists. They input large amount of information about their mental health status to these models. As a result, LLMs store, analyse, and memorise this data for enhancing output based on users needs and preferences.

In $2024$ a viral social media trend began emerging, during which users asked LLMs to draw a picture representing their life-style using the prompt ``Based on what you know about me, draw a picture of what my current life looks like.''~\cite{Mehta2024}. Being trained on users' data, LLMs could depict their current lives with great precision. 
While many users were fascinated about the LLMs capability in depicting their daily lives, cyber-aware users were even more alerted about how much data they are comfortable feeding these tools. This is another consequence many LLMs users never could imagine when they started feeding their personal information to these models on a daily basis.

\subsubsection{Case 4:} 
In early 2024, a group of researchers submitted an article~\cite{zhang2024three} to the ``Surfaces and Interfaces'' journal, which was first accepted and published, but then retracted due to unethical use of GenAI and self plagiarism. The authors in this paper have used GenAI to write the introduction section in this article. However, they have not acknowledged the use of GeAI and have not reviewed the output of the GenAI tool carefully enough to discard the initial communication the GenAI tool had with them stating that ``Certainly, here is a possible introduction for your topic''. Therefore, this article has been circulated over social media platforms by criticising researchers and academics since the authors neither acknowledged the assistance of GenAI, nor did they reviewed the output of GenAI, which is clearly evident in the first line of the introduction section in the retracted article. 

The other major concern with this article is self-plagiarism. Self-plagiarism is reusing work that has already been submitted for receiving credit in a course or has been published before. A careful review of~\cite{zhang2024three} reveals that Figure $1$ and Figure $2$ in the above mentioned article are duplicates of Figures in their previously published work~\cite{zhang2024performance}.

With advanced computer vision and text detection algorithms available these days, it will be easy for educational institutions, the research sector, and the publishing sector to detect replicated image and text content. This emphasises the importance of transparently acknowledging the use of GenAI tools in accordance with ethical guidelines. Otherwise, the consequences of unethical use of GenAI might forever impact the users' reputation and academic standing.


\subsection{Case Study of Benefits to Students and Educators}\label{Sec:Case Stud_Benefit}

Even though irresponsible and unethical use of GenAI tools in academic work cause academic and reputational harm, ethical and responsible use of these tools may sometimes provide some advantages to students.

In our case, we added an academic project writing assignments in the ``Industry Awareness Project'' course, during which  students are encouraged to ask ChatGPT to write a report on their topic with the same instructions (prompt) as given to students. The students are then asked to compare ChatGPT’s report on their topic with that of their own, and identify inaccuracies, bias, hallucinations, plagiarism, or any useful novel information in ChatGPT’s report. We found out that including this comparison task resulted in 1) decreasing the black-box use of LLMs in students' submissions,  2) enhancing students' critical analysis of AI generated content, and 3) increasing sense of confidence in their own academic capabilities.


\section{Guidelines for Ethical Use}\label{Sec:Ethics}

We encourage the students to consider a list of action points before and after applying GenAI tools in their academic work. 

\subsection{Ethical Guidelines Before Applying GenAI Tools}
It is essential to consider the following before using GenAI in collaborative academic report writing:\\

\noindent a) GenAI users should always ask their collaborators for their consent before applying these tools. Using GenAI tools without acquiring team members' consent is considered unethical.
\\

\noindent b) Check the guidelines and rules provided by the institution/venue to which  they are submitting their work. They can contact the editor in chief of the journal, the course coordinator at their institution, etc. to seek clarifications about the terms and conditions with regards to the use of GenAI prior to using these tools.
For example the Journal of Online Social Networks and Media\footnote{https://www.sciencedirect.com/journal/online-social-networks-and-media} requests the authors to declare the use of generative AI for manuscript writing even when used as a tool for enhancing the readability of small sections of their manuscript. This journal asks authors to declare that ``During the preparation of this work the author(s) used [NAME TOOL / SERVICE] in order to [REASON]. After using this tool/service, the author(s) reviewed and edited the content as needed and take(s) full responsibility for the content of the publication.''\\

\noindent c) Even if one has consent from their colleagues, clients, and organisation to use GenAI, they still should avoid feeding sensitive data to these tools.\\

\noindent d) Be aware of the latest bans on the use of GenAI tools. For example, DeepSeek~\cite{bi2024deepseek} has been banned from the Australian government devices due to national security and privacy concerns~\cite{BBC_DeepSeek_Ban}. Such bans are useful indicators of which tools we should completely avoid using.\\

\subsection{Ethical Guidelines After Applying GenAI Tools}

 Any type of GenAI application including generating text, image, editorial assistance, research, literature review, data curation, data analysis, and brainstorming must be acknowledged. 

\noindent As there are no universal clear guidelines about referencing GenAI content, and software tools, RMIT University advises students to reference text generated by AI according to section $7.2$ of the \textit{Fourth Edition of the Australian Guide to Legal Citation}\footnote{https://law.unimelb.edu.au/mulr/aglc/about$\#$fourth-edition}.~\cite{rmitAI2025} 
All students at RMIT University are advised to adhere to \href{https://rmit.libguides.com/referencing_AI_tools} {the guidelines published by RMIT University library} regarding proper referencing of AI-generated content.




Copying and pasting content generated by AI tools is not acceptable. This is mainly because no more than $10\%$ of academic reports should be made of direct quotes. Authors must always try to use their own words in academic writing. If they are unable to rewrite a phrase in their own terms while accurately maintaining its original sentiment and context, or if it is important to emphasise that a sentence or phrase was written or noted by a specific entity, they may use quotation marks. However, authors should be mindful of the $10\%$ limit.


If authors use GenAI tools for tasks such as brainstorming, conducting background research, shortlisting relevant literature, structuring their writing, paraphrasing sentences, editing grammar, or generating visual and aesthetic elements like diagrams and tables, it is crucial to acknowledge their use in the research method section or the body of the text. Authors should ensure they provide sufficient detail to inform the reader about the task, scale of use, name of the AI tool, the version, the date, and the prompt used. Such detailed information could be provided in the appendix to allow the reviewers to further investigate the accuracy of the content generated by AI model.

\section{Conclusion}

Writing a project report is much like solving mathematics problems or building muscles at the gym. Students will only develop analytical thinking, critical analysis, or formal writing skills when they invest time in reading new articles, engaging with diverse resources, and taking detailed notes. Establishing a routine to gather data, develop ideas, and organising the knowledge gained is essential. Building such skills requires both consistent effort and dedication. This is why educators encourage students to 
have a routine  for reading, writing, and discussion meetings to refine their ideas. Just as one cannot build muscles by merely observing their trainer doing exercise or learn mathematics by simply sitting through mathematics lectures, students cannot truly master their project topic by relying on GenAI to generate content for them. The regular process of researching, reading varied materials, writing down their learning, and discussing their ideas, both in their written report and during their regular group discussions, is key to mastering the writing of group projects.

Artificial `intelligence' models are inspired by the neural networks in the human brain. The way they are trained  is similar to how the human brain learns new concepts. Among the strongest deep learning and generative modelling techniques are reinforcement learning (RL)~\cite{sutton1998reinforcement} and ProGAN~\cite{karras2017progressive}. RL models are trained through interacting with the environment, taking actions, making mistakes, receiving feedback, and improving their performance based on the feedback (reward) over time. ProGAN is trained to generate new data by slowly and continuously progressing toward the objective. The intuition behind ProGAN is that you must \textit{progress slowly and gradually} in learning a concepts with \textit{depth and quality}. This is to say that, even the advanced deep learning techniques learn through making mistakes during gradual training, and by receiving constant feedback. Thus, to master their project topic, students need to commit to learning consistently and not be afraid of making mistakes. Making mistakes throughout the learning process is an effective learning mechanism which enables critical thinking.

It is important for the students to be confident of their abilities, and to acknowledge that they can do better than a trained machine learning algorithm when it comes to novel ideas, and critical unbiased analysis. Students are encouraged to consider their projects as an opportunity to develop and improve their writing skills while they have access to resources that can support them during this process.

In summary, this article aims to advise students about the risks and circumstances of using GenAI, particularly in collaborative academic report writing. We encourage students to take into consideration the risk-benefit analysis of using GenAI models, and to educate themselves about the terms and conditions of these tools before feeding them with data. Last but not least, ethical research conduct dictates that any use of GenAI should be transparently discussed with all team members and acknowledged before submitting the work.

\bibliographystyle{IEEEtran}
\input{main.bbl}

\end{document}